\documentclass[sn-mathphys-num]{sn-jnl}%

\usepackage{amssymb}
\usepackage{amsmath}

\usepackage{lineno}
\usepackage{eurosym}
\usepackage[ruled,vlined]{algorithm2e}
\usepackage{subfig}
\usepackage{url}
\usepackage{xcolor}
\usepackage[inline,shortlabels]{enumitem}

\usepackage{natbib}
\usepackage{graphicx}%
\usepackage{multirow}%
\usepackage{amsmath,amssymb,amsfonts}%
\usepackage{amsthm}%
\usepackage{mathrsfs}%
\usepackage[title]{appendix}%
\usepackage{xcolor}%
\usepackage{textcomp}%
\usepackage{manyfoot}%
\usepackage{booktabs}%
\usepackage{listings}%
\usepackage[round-mode=places,round-precision=2,round-pad=false,group-separator={,},output-decimal-marker={.}]{siunitx}

\newcommand{\bigO}{\mathcal{O}}

\newcommand{\mathsc}[1]{{\normalfont\textsc{#1}}}

\newcommand{\partitions}{p}
\newcommand{\partitionsPerDim}{m}
\newcommand{\cores}{c}
\newcommand{\representatives}{rep}

\newcommand{\sky}{\mathsc{Sky}}

\newcommand{\grid}{\mathsc{Grid}\xspace}
\newcommand{\angular}{\mathsc{Angular}\xspace}
\newcommand{\sliced}{\mathsc{Sliced}\xspace}

\newcommand{\none}{\mathsc{None}\xspace}

\newcommand{\gridres}{\mathsf{gres}}

\newcommand{\gridproj}{\mathsf{gproj}}

\newcommand{\dominates}{\prec}

\newcommand{\dimensions}{d}

\newcommand{\positivereals}{\mathbb{R^+}}

\newcommand{\positivenaturals}{\mathbb{N^+}}

\newcommand{\size}{N}

\newcommand{\uniform}{\texttt{UNI}\xspace}
\newcommand{\anticorrelated}{\texttt{ANT}\xspace}
\newcommand{\nba}{\texttt{NBA}\xspace}
\newcommand{\household}{\texttt{HOU}\xspace}
\newcommand{\emp}{\texttt{EMP}\xspace}
\newcommand{\sensors}{\texttt{SEN}\xspace}
\newcommand{\zillow}{\texttt{RES}\xspace}

\newcommand{\griddominates}{\prec_G}

\newcommand{\mycomment}[1]{}

\newtheorem{definition}{Definition}

\newcommand{\logSep}{\,.\,\,}

\def\codeif{\mbox{\upshape\textbf{if}}}
\def\codethen{\mbox{\upshape\textbf{then}}}

\def\codeforeach{\mbox{\upshape\textbf{for each}}}
\def\codedo{\mbox{\upshape\textbf{do}}}
\def\codein{\mbox{\upshape\textbf{in}}}

\def\codereturn{\mbox{\upshape\textbf{return}}}

\def\codenil{\mbox{\upshape\textbf{nil}}}

\begin{document}

\title{Parallelizing the Computation of Robustness for Measuring the Strength of Tuples}

\author*[1]{\fnm{Davide} \sur{Martinenghi}}\email{davide.martinenghi@polimi.it}

\affil*[1]{\orgdiv{DEIB}, \orgname{Politecnico di Milano}, \orgaddress{\street{Piazza Leonardo 32}, \city{Milan}, \postcode{20133}, \country{Italy}}}

\abstract{
Several indicators have been recently proposed for measuring various characteristics of the tuples of a dataset -- particularly, the so-called \emph{skyline} tuples, i.e., those that are not dominated by other tuples.
Numeric indicators are very important as they may, e.g., provide an additional criterion to be used to rank skyline tuples and focus on a subset thereof.
We concentrate on an indicator of robustness that may be measured for any skyline tuple $t$: grid resistance, i.e.,
how large value perturbations can be tolerated for $t$ to remain non-dominated (and thus in the skyline).

The computation of this indicator typically involves one or more rounds of computation of the skyline itself or, at least, of dominance relationships. Building on recent advances in partitioning strategies allowing a parallel computation of skylines, we discuss how these strategies can be adapted to the computation of the indicator.
}


\maketitle

\section{Introduction}
\label{intro}

Multi-Criteria Analysis aims to identify the most suitable alternatives in datasets characterized by multiple attributes. This challenge is prevalent in many data-intensive fields and has been amplified by the advent of Big Data, which emphasizes the importance of efficiently searching through vast datasets.

Skyline queries are a widely used method to address this issue, filtering out alternatives that are dominated by others. An alternative $a$ is said to dominate $b$ if $a$ is at least as good as $b$ in all attributes and strictly better in at least one. Non-dominated alternatives are valuable because they represent the top choice for at least one ranking function, thus offering a comprehensive view of the best options.

A common limitation of skyline queries is their complexity, generally quadratic in the dataset size, which poses challenges in Big Data contexts. To mitigate this, researchers have been exploring dataset partitioning strategies to enable parallel processing, thereby reducing overall computation time.
The typical approach involves a two-phase process: first, computing local skylines within each partition, and second, merging these local skylines to create a pruned dataset for a final skyline computation. The aim is to eliminate as many dominated alternatives as possible during the local skyline phase, minimizing the dataset size for the final computation.

Another limitation of skylines as a query tool is that they may return result sets that are too large and thus of little use to the final user. A way around this problem is to equip skyline tuples with additional numeric indicators that measure their ``strength'' according to various characteristics. With this, skyline tuples can be ranked and selected accordingly to offer a more concise result to the final user.
Several previous research attempts
 have proposed a plethora of such indicators.
We focus here on an indicator,
 called grid resistance, measures how robust a skyline tuple is to slight perturbations of its attribute values.
Quantizing tuple values (e.g., in a grid) affects dominance, since, as the quantization step size grows, more values tend to collapse, causing new dominance relationships to occur.

The computation of this indicator may involve, in turn, several rounds of computation of the skyline or of dominance relationships. In this respect, the parallelization techniques that have been developed for computing skylines may prove especially useful for the indicators, too. 
In this work, we describe the main parallelization opportunities for computing 
the grid resistance of skyline tuples, and provide an experimental evaluation that analyzes the impact of such techniques in several practical scenarios.

\section{Preliminaries}
\label{sec:prelim}

We refer to datasets consisting of numeric attributes. Without loss of generality, the domain we consider is the set of non-negative real numbers $\positivereals$. A schema $S$ is a set of attributes $\{A_1,\ldots, A_\dimensions\}$, and a tuple $t=\langle v_1,\ldots,v_\dimensions\rangle$ over $S$ is a function associating each attribute $A_i\in S$ with a value $v_i$, also denoted $t[A_i]$, in $\positivereals$; a relation over $S$ is a set of tuples over $S$.

A \emph{skyline} query~\cite{DBLP:conf/icde/BorzsonyiKS01}  takes a relation $r$ as input and returns the set of tuples in $r$ that are dominated by no other tuples in $r$, where dominance is defined as follows.

\begin{definition}
Let $t$ and $s$ be tuples over a schema $S$; $t$ \emph{dominates} $s$, denoted $t\dominates s$, if, for every attribute $A\in S$, $t[A]\leq s[A]$ holds and there exists an attribute $A' \in S$ such that $t[A']<s[A']$ holds.
The \emph{skyline} $\sky(r)$ of a relation $r$ over $S$ is the set $\{t \in r \mid \nexists s\in r \logSep s\dominates t\}$.
\end{definition}

We shall consider attributes such as ``cost'', where smaller values are preferable; however, the opposite convention would also be possible.

A tuple can be associated with a numeric score via a \emph{scoring function} applied to the tuple's attribute values. For a tuple $t$ over a schema $S=\{A_1,\ldots, A_\dimensions\}$, a scoring function $f$ returns a score
$f(t[A_1],\ldots, t[A_\dimensions])\in \positivereals$,
also indicated $f(t)$.
As for attribute values, we set our preference for lower scores (but the opposite convention would also be possible).

Although skyline tuples are unranked, they can be associated with extra numeric values by computing appropriate indicators and ranked accordingly. To this end, we consider a robustness indicator.

The indicator called \emph{grid resistance}, and denoted $\gridres(t;r)$, measures how robust skyline tuple $t$ is with respect to a perturbation of the attribute values of the tuples in $r$, i.e., whether $t$ would remain in the skyline.
In~\cite{CM:PACMMOD2024}, this is done by snapping tuple values to a grid divided in $g$ equal-size intervals in each dimension: the more a skyline tuple resists to larger intervals, the more it is robust.
The \emph{grid-projection} $\gridproj(t,g)$ of $t$ on the grid is defined as
\(
\gridproj(t,g)=\left\langle \frac{\lfloor t[1]\cdot g\rfloor}{g}, \ldots, \frac{\lfloor t[\dimensions]\cdot g\rfloor}{g} \right\rangle,
\)
and corresponds to the lowest-value corner of the cell that contains $t$.
When tuples are mapped to their grid-projections, we obtain a new relation $\gridproj(r,g)=\{\gridproj(t,g)\mid t\in r\}$, in which new dominance relationships may occur.
The \emph{grid resistance} $\gridres(t;r)$ of $t$ is the smallest value of $g^{-1}$ for which $t$ is no longer in the skyline.
\begin{definition}
Let $r$ be a relation and $t\in\sky(r)$. The grid resistance $\gridres(t;r)$ of $t$ in $r$ is $\underset{g\in\positivenaturals} {\min} \{g^{-1}\mid \gridproj(t,g)\notin\sky(\gridproj(r,g))\}$. We set $\gridres(t;r)=1$ if $t$ never exits the skyline.

\end{definition}

\section{Parallel Algorithms}
\label{sec:parallel}

In this section, we first present the main partitioning strategies adopted in the literature, and then describe how they can be adapted for computing the indicators.

Such strategies have been developed assuming a general scheme for parallelizing the computation of the skyline that consists of the following phases:
\begin{enumerate}
	\item each partition is processed independently and in parallel to produce a ``local'' skyline; the union of these local skylines may still contain tuples that are dominated by tuples in other partitions;
	\item the final result is obtained by applying the skyline operator to the union of all the local skylines by removing all residual dominated tuples.
\end{enumerate}
The input to the first phase may also include additional meta-information that will accelerate the process. The last phase is typically executed sequentially, but there are ways to parallelize this phase, too~\cite{ciaccia2024optimizationstrategiesparallelcomputation}.

\subsection{Partitioning strategies}

We now review three of the main partitioning strategies available in the literature: grid partitioning, angle-based partitioning, and sliced partitioning.

\begin{figure}%
\centering
\subfloat[][{\grid}]
{\includegraphics[width=0.3\textwidth]{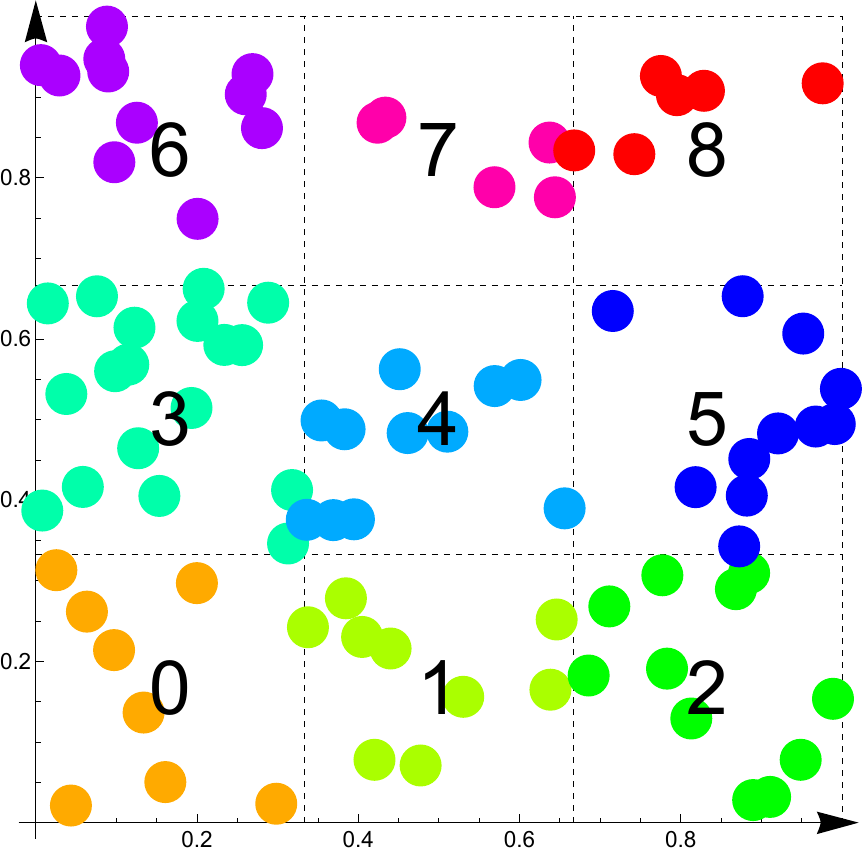}\label{fig:gridPartitioning}}%
\quad
\subfloat[][{\angular}]
{\includegraphics[width=0.3\textwidth]{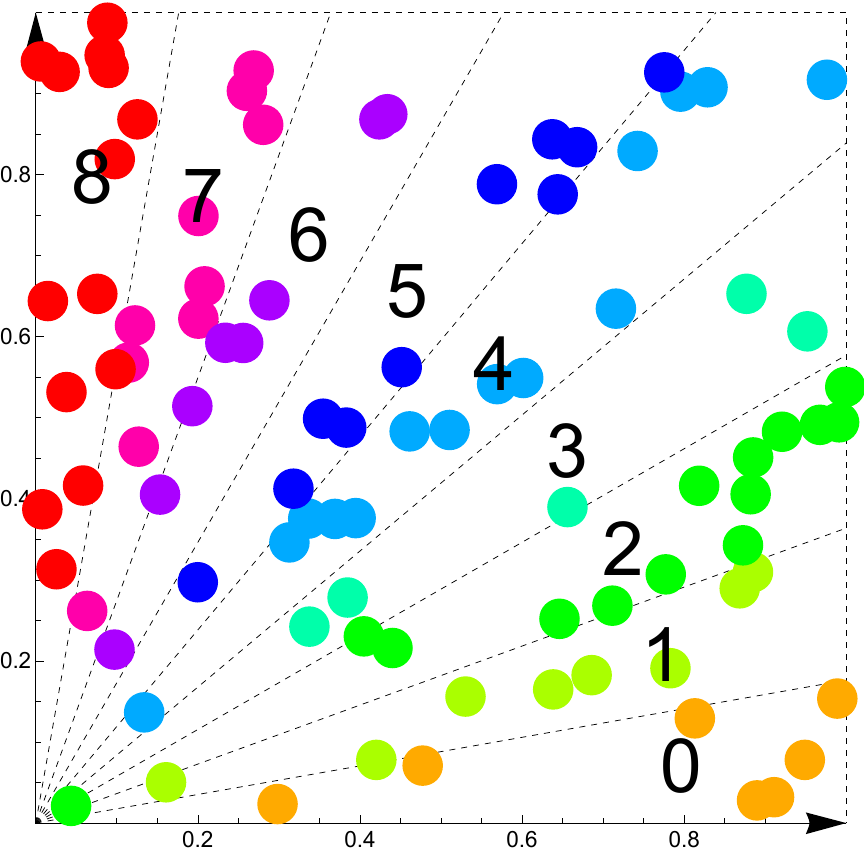}\label{fig:anglePartitioning}}%
\quad
\subfloat[][{\sliced}]
{\includegraphics[width=0.3\textwidth]{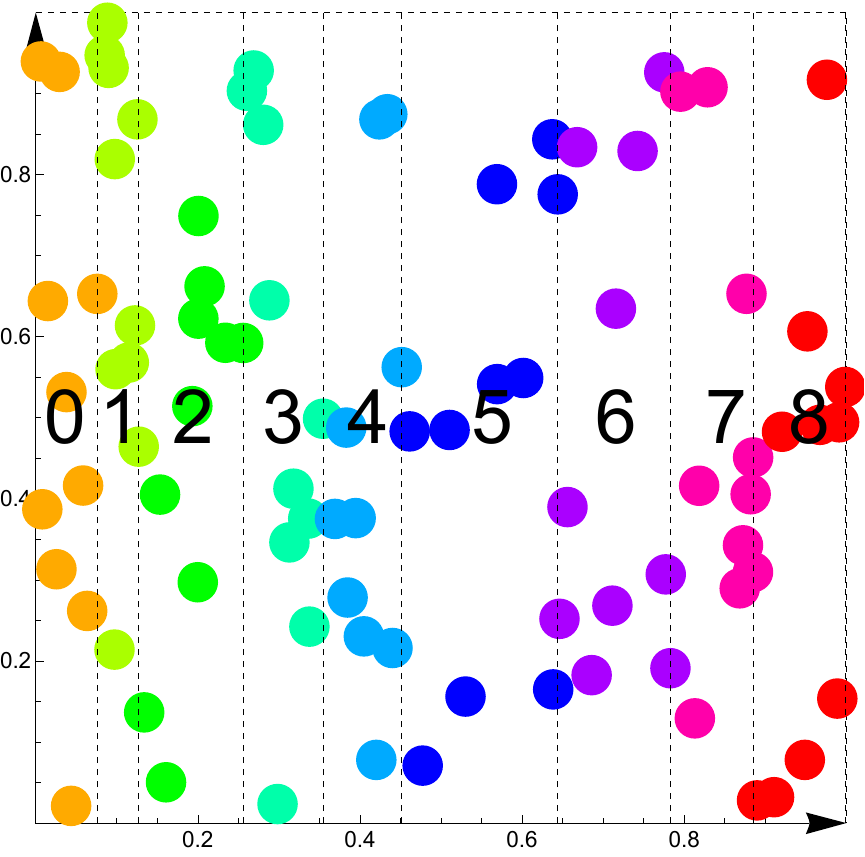}\label{fig:slicedPartitioning}}%
\caption{Partitioning Strategies illustrated on a uniform dataset.}\label{fig:strategies}%
\end{figure}

Figure~\ref{fig:strategies} illustrates the different partitioning strategies as applied to a uniformly distributed dataset of 90 tuples with 9 partitions, where different partitions are represented with different colors.

\smallskip

\emph{Grid Partitioning}~\cite{DBLP:conf/edbt/MullesgaardPLZ14} (\grid) partitions the space into a grid of equally sized cells, resulting in a total of $\partitions=\partitionsPerDim^\dimensions$ partitions, where $\dimensions$ denotes the total number of dimensions and $\partitionsPerDim$ the number of slices in which each dimension is split.
This strategy additionally entails a dominance relationship applied to grid cells (and thus to partitions), which allows us to avoid processing certain partitions completely.

We identify a given cell $c_i$ with its \emph{grid coordinates}, $\langle c_i[1],\ldots,c_i[\dimensions]\rangle$, with $1\leq c_i[j]\leq m$, $j=1,\ldots,\dimensions$. With this, we can introduce grid dominance.

\begin{definition}[Grid dominance]
For grid cells $c_i$ and $c_h$, $c_i$ \emph{grid-dominates} $c_h$, denoted $c_i \griddominates c_h$, if for every dimension $j$, $j=1,\ldots,\dimensions$, we have $c_i[j] < c_h[j]$.  
\end{definition}

If $c_i$ grid-dominates $c_h$ then all tuples in $c_i$ dominate all tuples in $c_j$, so if $c_i$ is not empty, $c_j$ can be disregarded altogether for computing the skyline.

Assigning a partition number, shown in Figure~\ref{fig:gridPartitioning} for grid partitioning, to a tuple $t$ can be done as follows, assuming, for simplicity, all values to be in $[0,1)$:
\[
\partitions(t) = \sum_{i=1}^{\dimensions} \lfloor t[A_i]\cdot \partitionsPerDim \rfloor \cdot \partitionsPerDim^{i-1}
\]
\noindent where $A_i$ is the $i$-th attribute. As an example of grid-dominance that can be seen in the figure, partition 1 grid-dominates partitions 5 and 8.

\smallskip

\emph{Angle-based Partitioning}~\cite{DBLP:conf/sigmod/VlachouDK08} (henceforth: \angular) partitions the space wrt angular coordinates, after converting Cartesian to hyper-spherical coordinates, which provide a better workload balance across partitions than with \grid.

The partition number (shown in Figure~\ref{fig:anglePartitioning} for \angular) is computed for every tuple $t$ based on hyper-spherical coordinates, including a radial coordinate $r$ and $\dimensions-1$ angular coordinates $\varphi_1, \ldots, \varphi_{\dimensions-1}$, obtained through standard geometric considerations from the Cartesian coordinates.
The partition number of $t$
is then computed as follows:
\begin{equation}\label{eq:index-angular}
\partitions(t)=\sum_{i=1}^{\dimensions-1}\left\lfloor \frac{2\varphi_i}{\pi}\partitionsPerDim \right\rfloor
\partitionsPerDim^{i-1}
\end{equation}
where $\partitionsPerDim$ is, again, the number of slices in which each (angular) dimension is divided, which essentially amounts to grid partitioning on angular coordinates.

\smallskip

\emph{Sliced Partitioning}~\cite{ciaccia2024optimizationstrategiesparallelcomputation} (\sliced) first sorts the dataset with respect to one chosen dimension, then (unlike \grid and \angular) determines any given number $p$ of equi-numerous partitions.
The partition number (shown in Figure~\ref{fig:slicedPartitioning} for \sliced) of the $i$-th tuple $t$ in the ordering is simply computed as follows:
\[
\partitions(t) = \left\lfloor \frac{(i-1)\cdot \partitions}{\size}\right\rfloor,
\]
where $\size$ is the number of tuples in the dataset.

\medskip

We observe that all the partitioning strategies can be improved by resorting to several optimization opportunities.
We reconsider here the so-called 
\emph{Representative Filtering}, 
as presented in~\cite{ciaccia2024optimizationstrategiesparallelcomputation}.
Representative Filtering consists in pre-computing a few potentially ``strong'' tuples to be shared across all partitions, since they may have a high potential for dominating other tuples, thus further removing redundant tuples in the local skylines of each partition.
A simple technique to select representative tuples consists in choosing the top-$k$ results according to any given monotone scoring function $f$ of the dataset's attributes. This can be done, e.g., in $\bigO(\size \log k)$ by using a (max-)heap as follows:
\begin{enumerate*}[label=\emph{(\roman*)}]
	\item insert the first $k$ tuples in the heap;
	\item scan the rest of the dataset and, for each tuple $t$, if $f(t) < f(t_k)$ (where $t_k$ is the tuple at the root of the heap) replace $t_k$ with $t$ and re-adjust (heapify) the heap.
\end{enumerate*}
One can even obtain the same result in $\bigO(\size + k\log k)$ by 
\begin{enumerate*}[label=\emph{(\roman*)}]
\item executing a selection algorithm running in $\bigO(\size)$ to find the $k$-th smallest tuple according to $f$,
\item using this as a pivot in the \texttt{QuickSort} sense to separate the $k$ smallest tuples from the others, and
\item finally sorting the $k$ tuples in $\bigO(k \log k)$.
\end{enumerate*}
See, e.g.,~\cite{DBLP:books/daglib/0023376} for details about the selection algorithm.
Note that, after selecting the $k$ representatives, those that are dominated by other representatives should be discarded, since they do not add any pruning power to the set; this means that the actual number of tuples used for the subsequent filtering is potentially lower than $k$.

\subsection{Computing the indicator}

Finding $\gridres$ requires recomputing dominance on grid-projected datasets for various values of the grid interval $g$. Luckily, the $\gridres$ operator is stable, i.e., it does not depend on dominated tuples, and therefore we can focus on skyline tuples alone.
Algorithm~\ref{alg:gres} shows the pseudo-code illustrating the idea.
The grid interval varies between $2$ (smallest meaningful value) and an upper bound $\bar{g}$ that depends on the dataset. In particular, we have the guarantee that no new dominance relationship may occur when $g>\bar{g}=\lfloor\ell^{-1}\rfloor$, where $\ell$ is the absolute value of the smallest non-zero difference on the same attribute between any two tuples (line~\ref{line:gres-g-bar}).
It suffices then to compute, for each value of $g$ among $\bar{g}, \bar{g}-1, \ldots,2$ (line~\ref{line:gres-for-each-g}),
the skyline $\sky(\gridproj(r,g))$ (line~\ref{line:gres-skyline})
and, for each tuple $t\in\sky(r)$, test whether
$\gridproj(t,g)\in\sky(\gridproj(r,g))$; the inverse of the first value of $g$ for which membership does not hold is $\gridres(t,r)$  (lines~\ref{line:gres-for-each-t}-\ref{line:gres-set-map}).
If we ignore the dependence on $\bar{g}$, which is dataset-dependent, the complexity of computing $\gridres$ for a given tuple is $\bigO(S^2)$, where $S=|\sky(r)|\in\bigO(\size)$, i.e., in the worst case $\bigO(\size^2)$.
It is strikingly evident that computing $\gridres$ requires several rounds of computation of the skyline (although on a potentially much smaller dataset than the starting one, since dominated tuples can be disregarded completely). In this respect, adopting the partitioning strategies that typically quicken skyline computation may be beneficial for $\gridres$, too.

\begin{algorithm}[t]
   \scalebox{.82}
   {
    \begin{minipage}{1.33\textwidth}
	\begin{enumerate}
	   \item[Input:] \emph{skyline $s=\sky(r)$}
	   \item[Output:] \emph{a map from every tuple $t\in s$ to $\gridres(t,r)$}
		\item\label{line:gres-map} $map := \emptyset$  // \textit{the result map, initially empty}
		\item\label{line:gres-g-bar} $\bar{g} := \lfloor\ell^{-1}\rfloor$  // \textit{where $\ell$ is the minimum possible value for $\gridres$}
	   \item\label{line:gres-for-each-g} \codeforeach\ $g$ \codein\ $\bar{g},\ldots,2$ \codedo
	   \item\label{line:gres-skyline} \quad $s' := \sky(\gridproj(r,g))$
	   \item\label{line:gres-for-each-t} \quad \codeforeach\ $t$ \codein\ $s$ \codedo
		\item\label{line:gres-set-map} \quad \quad \codeif\ $\gridproj(t,g)\notin s' \land map(t) = \codenil$ \codethen\ $map(t) := g^{-1}$
      \item\label{line:gres-for-each-t-afterwork} \codeforeach\ $t$ \codein\ $s$ \codedo
      \item\label{line:gres-never-exited} \quad \codeif\ $map(t) = \codenil$ \codethen\ $map(t) := 1$  // \textit{$t$ never exited the skyline}
      \item\label{line:gres-return} \codereturn\ $map$
	\end{enumerate}	    
	\end{minipage}
   }
	\caption{Algorithmic pattern for computing $\gridres$.}
	\label{alg:gres}
\end{algorithm}

\section{Experiments}
\label{sec:experiments}

In this section, we test effectiveness and efficiency of the proposed algorithmic pattern (Algorithm~\ref{alg:gres}) on a number of scenarios.

The datasets we use for the experiments comprise both synthetic and real datasets.
For synthetic datasets, we produce, for any value of $\dimensions$ and $\size$ mentioned in Table~\ref{tab:operating_parameters}, two $\dimensions$-dimensional datasets of size $\size$ with values in the $[0,1)$ interval: one with values anti-correlated across different dimensions ($\anticorrelated$) and one with uniformly distributed values ($\uniform$). For each dataset, we generate 5 different instances; our results report averages over those instances.

The real datasets we adopt are the result of a cleaning, normalization and attribute selection process that resulted in the following:
\begin{itemize}
\item $\nba$ -- all-time stats for $\num{4832}$ NBA players from \url{nba.com} as of Oct. 2023, from which we retained 2 attributes;
\item $\household$ -- \num{127931} 6D tuples regarding household data scraped from \url{www.ipums.org};
\item $\emp$ -- $\num{291825}$ 6D tuples about City employees in San Francisco~\cite{emp};
\item $\zillow$ -- real estate data from \url{zillow.com}, with $\num{3569678}$ 6D tuples;
\item $\sensors$ -- sensor data with 7 numeric attributes and $\num{2049280}$ tuples~\cite{sensors}.
\end{itemize}
In our experiments, we vary several parameters (shown in Table~\ref{tab:operating_parameters}) to measure their impact, on various datasets, on the number of dominance tests required to compute the final result:
\begin{itemize} 
    \item dataset cardinality $\size$,
    \item number of dimensions $\dimensions$, and
    \item number of partitions $\partitions$.
\end{itemize}
Since with \grid and \angular, not all values of $\partitions$ are possible, we run them with a value of $\partitions$ that is closest to the target number shown in Table~\ref{tab:operating_parameters}, provided that the number of resulting partitions is greater than $1$.

The number of dominance tests incurred during the various phases of our algorithmic patterns provides us with an objective measure of the effort required for computing the indicators, and this independently of the underlying hardware configuration.

Any computing infrastructure will essentially face \emph{(i)} an overhead for the parallel phase, \emph{(ii)} one for the sequential phase, and \emph{(iii)} one for orchestration of the execution and communication between nodes. While \emph{(iii)} depends on the particular infrastructure, \emph{(i)} and \emph{(ii)} will depend directly on the number of dominance tests. In particular, the cost of the sequential phase will be proportional to the number of dominance tests made during that phase, while, in the parallel phase, the cost will be proportional to the largest amount of dominance tests that need to be pipelined by the parallel computation.
Put simply, if there is at least one core per partition, the parallel phase will roughly cost as much as the processing cost incurred by the heaviest partition; if the partitions/cores ratio is $k$, then the cost of the parallel phase will be scaled by a factor $k$.

Besides the objective measure given by dominance tests, we also measure execution times by parallelizing the tasks on a machine sporting an Apple chip with 16 cores,
 and controlling the number of active cores through semaphores.

\begin{table}[h]
   \centering
   \caption{Operating parameters for testing efficiency (defaults in bold).}
       \begin{tabular}{|l|l|}
           \hline
               Full name                           & Tested value \\
           \hline
               Distribution                        & synthetic: \anticorrelated, \uniform; real: \nba, \household, \emp, \zillow, \sensors \\
               Synthetic dataset size ($\size$)                & 100K, 500K, \textbf{1M}, 5M, 10M \\
               \# of dimensions ($\dimensions$)    & 2, \textbf{3}, 4, 5, 6, 7 \\
               \# of partitions ($\partitions$)    &  \textbf{16}, 32, 64, 128\\
               \# of representatives ($\representatives$)    &  \textbf{0}, 1, 10, 100, 1000\\
               \# of cores ($\cores$)    & 2, 4, 8, \textbf{16} \\
           \hline
       \end{tabular}
   \label{tab:operating_parameters}
\end{table}

We now report our experiments on the computation of $\gridres$ with the different partitioning strategies.
Before starting the experiments, we observe that, while an exact determination of $\gridres$ would require determining $\ell$ as in line~\ref{line:gres-g-bar} of Algorithm~\ref{alg:gres} and its inverse $\bar{g}$, the actual value of $\ell$ may be impractically small.
Bearing in mind the aim of the $\gridres$ indicator was to determine the tuples that are ``strong'' wrt grid resistance, and that, for very small values of $\ell$, the corresponding value of $\gridres$ would be insignificant, we chose to move to a more practical option.
Therefore, instead of looking for the smallest non-zero difference (in absolute value) between any two values on the same attribute in the dataset, we simply set $\bar{g}=25$ as a reasonable threshold of significance for the number of grid intervals.

\begin{figure}%
\centering
\subfloat[][{\anticorrelated}]
{\includegraphics[width=0.49\textwidth]{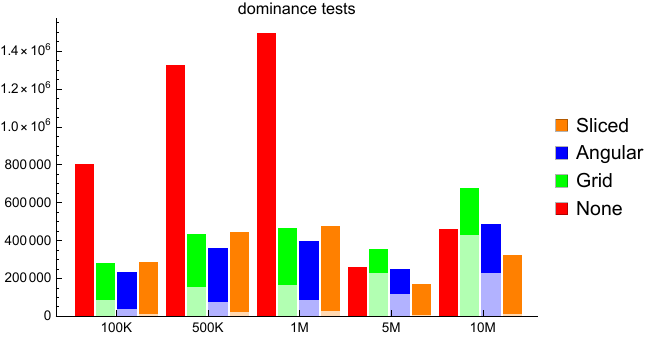}\label{fig:ant-varyingSize-p16}}%
\subfloat[][{\uniform}]
{\includegraphics[width=0.49\textwidth]{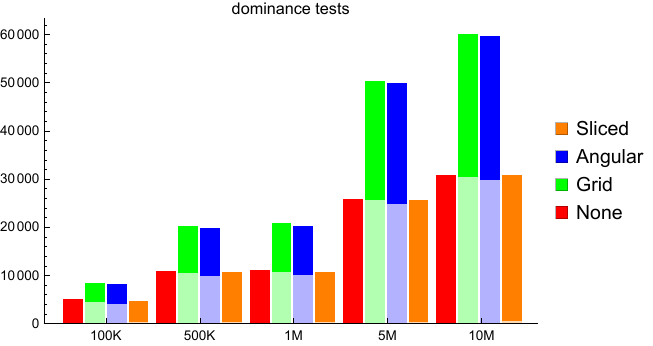}\label{fig:uni-varyingSize-p16}}%
\caption{Number of dominance tests incurred by the various partitioning strategies with a default number of partitions ($\partitions=16$) and varying dataset sizes on \anticorrelated (\ref{fig:ant-varyingSize-p16}) and \uniform (\ref{fig:uni-varyingSize-p16}).}\label{fig:varyingSize-p16}%
\end{figure}

\medskip

\noindent\textbf{Varying the dataset size $\size$.}
Our first experiment focuses on the effect of the dataset size on the number of dominance tests needed to compute the result, while keeping all the other operating parameters to their default values as indicated in Table~\ref{tab:operating_parameters}.
Figure~\ref{fig:ant-varyingSize-p16} focuses on \anticorrelated and reports stacked bars for each of the partitioning strategies, in which the lower part refers to the largest number of dominance tests made in any partition during the parallel phase, while the top part indicates the number of dominance tests made during the final phase.
We observe that up to $\size=1$M, all partitioning strategies are beneficial with respect to no partitioning (indicated as \none), with \angular as the most effective strategy and \sliced as the strategy with the lowest parallel costs, due to an ideal balancing of the number of tuples in each partition.
For larger sizes, however, \grid becomes less effective, \angular is on a par with \none and \sliced becomes the most effective strategy.
This is due to a sort of non-monotonic behavior with respect to the number of skyline points: for a growing dataset size, the size of the skyline typically also grows; however, when we move from $\size = 1$M to $\size = 5$M there is a fall from $1022$ to only $527$ tuples in $|\sky|$. This is due to the fact that the larger dataset, besides containing more tuples, also includes some very strong tuples that dominate most of the remaining ones. A similar effect is observed for $\size = 10$M, for which the skyline size only grows to $650$ tuples, and is therefore smaller than with $\size=1$M, and actually even smaller than with $\size = 500$K (where $|\sky|=940$) $\size = 100$K (where $|\sky|=719$). These numbers refer to one of our exemplar instances, but similar behaviors, which are due to the way in which data are generated (we adopted the widely used synthetic data generator proposed by the authors of~\cite{DBLP:conf/icde/BorzsonyiKS01}), are nonetheless common to all 5 repetitions of the experiments that we performed.

Figure~\ref{fig:uni-varyingSize-p16} shows the stacked bars for the \uniform dataset. Here, the benefits of parallelization are essentially lost, at least for \grid and \angular, due to the extremely small skyline sizes that occur with uniform distributions (varying from $75$ tuples for $\size=100$K to 101 for $\size=10$M). The \sliced partitioning strategy manages to still offer slight improvements with respect to \none by doing very little removal work during the parallel phase, and therefore a final phase only slightly lighter than with \none.

\begin{figure}%
\centering
\subfloat[][{\anticorrelated}]
{\includegraphics[width=0.49\textwidth]{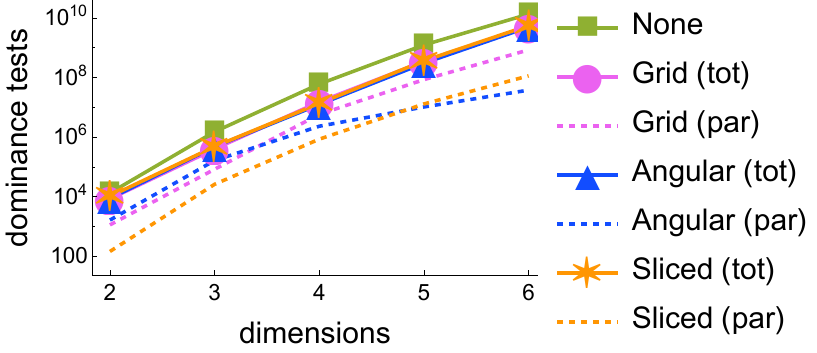}\label{fig:ant-varyingD-p16}}%
\subfloat[][{\uniform}]
{\includegraphics[width=0.49\textwidth]{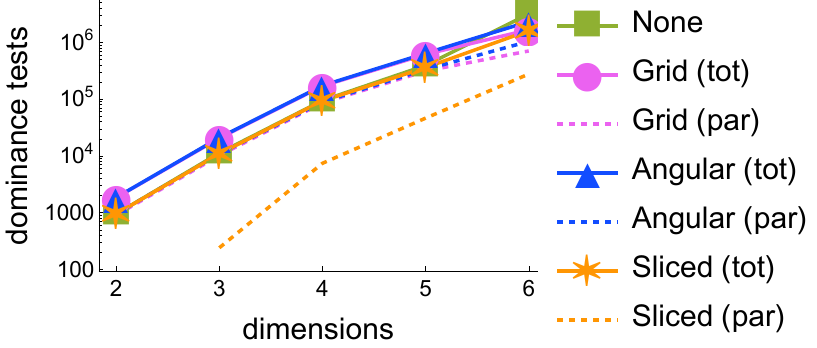}\label{fig:uni-varyingD-p16}}%
\caption{Number of dominance tests with a default number of partitions ($\partitions=16$) as the number of dimensions varies on \anticorrelated (\ref{fig:ant-varyingD-p16}) and \uniform (\ref{fig:uni-varyingD-p16}) datasets with $\size=1$M tuples.}\label{fig:varyingD-p16}
\end{figure}

\medskip

\noindent\textbf{Varying the number of dimensions $\dimensions$.}
Figure~\ref{fig:varyingD-p16} shows how the number of dominance tests varies as the number of dimensions in a synthetic dataset grows.
The plots need to use a logarithmic scale since the number of tests grows exponentially, as an effect of the ``curse of dimensionality''.
With the \anticorrelated datasets (Figure~\ref{fig:ant-varyingD-p16}), all partitioning strategies offer significant gains with respect to the plain sequential execution, with savings of up to nearly 80\%, and never under 60\% for $\dimensions > 2$. Indeed, when $\dimensions = 2$ the skyline consists of only 62 tuples, so the benefits of partitioning are smaller. In particular, \grid is the top performer for $\dimensions \leq 4$, while \angular wins for $\dimensions \geq 5$, although the differences between the best and the worst strategies are always under 5\%.
The dashed lines indicate the largest number of dominance tests made in any partition during the parallel phase, while the solid lines indicate the overall cost of a parallel execution, i.e., by adding to the previous component the number of dominance tests of the final phase.

In the case of \uniform datasets, skyline sizes are very small for low $\dimensions$, with as few as 11 tuples when $\dimensions=2$. Clearly, parallelizing does not pay off in such circumstances. For larger values of $\dimensions$ the gains are more significant, reaching nearly 50\% when $\dimensions = 6$ with the \sliced strategy, which proves to be the most adequate for this kind of dataset in all scenarios, while \grid and \angular reach 26\% and 47\%, respectively, under the same conditions.
We also note that, when there are more partitions than skyline points (as is the case, e.g., with $\dimensions=2$, in which $p=16>11=|\sky|$), the partitions in \sliced will have either 0 or 1 points, so no dominance test will ever take place in the parallel phase, which is then completely ineffective.

\begin{figure}%
\centering
\subfloat[][{\anticorrelated}]
{\includegraphics[width=0.49\textwidth]{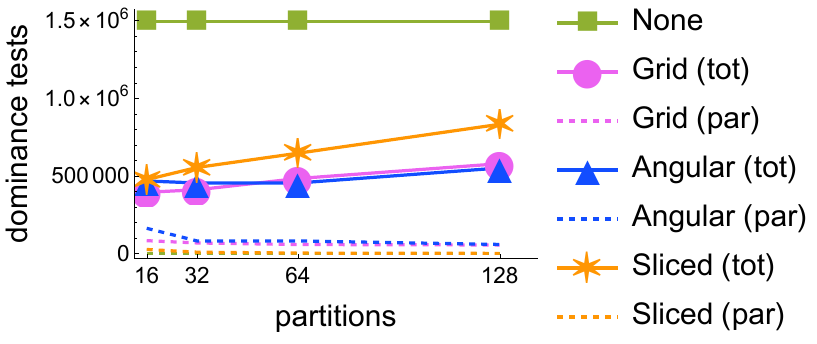}\label{fig:ant-varying-partitions}}%
\subfloat[][{\uniform}]
{\includegraphics[width=0.49\textwidth]{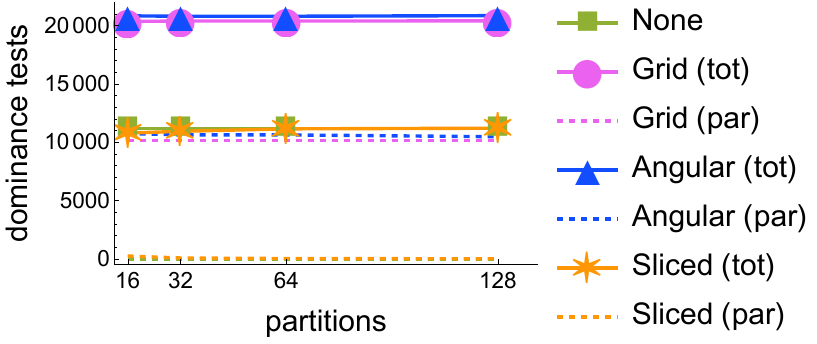}\label{fig:uni-varying-partitions}}%
\caption{Number of dominance tests as the number of partitions varies on \anticorrelated (\ref{fig:ant-varying-partitions}) and \uniform (\ref{fig:uni-varying-partitions}) 3D datasets with $\size=1$M tuples.}\label{fig:varying-partitions}
\end{figure}

\medskip

\noindent\textbf{Varying the number of partitions $\partitions$.}
Figure~\ref{fig:varying-partitions} shows the effect of the number of partitions on the number of dominance tests.
While the partitioning is always beneficial with the \anticorrelated datasets (Figure~\ref{fig:ant-varying-partitions}), increasing the number of partitions only increases the overhead.
This phenomenon is due to two causes: the first one is the relatively small size of the input dataset for computing $\gridres$, i.e., the size of the skyline of a 3D dataset of $1$M tuples (1022 tuples in our case), which makes it less open to more intense parallelization opportunities; the second reason is that, since the input consists of skyline tuples of the original dataset, for small grid intervals (i.e., for larger values of $g$ as used in Algorithm~\ref{alg:gres}), the grid projections of these skyline tuples will almost never exit the skyline, so that dominance tests will be ineffective and the final phase will be predominant, as can be clearly seen, e.g., in Figure~\ref{fig:ant-varying-partitions}.
For \uniform, these effects do not change, but are less visible because of the much smaller skyline size involved (just 78 tuples), which makes \grid and \angular completely ineffective, while \sliced maintains performances on a par with or slightly better than \none.

\begin{figure}%
\centering
\subfloat[][{\anticorrelated}]
{\includegraphics[width=0.49\textwidth]{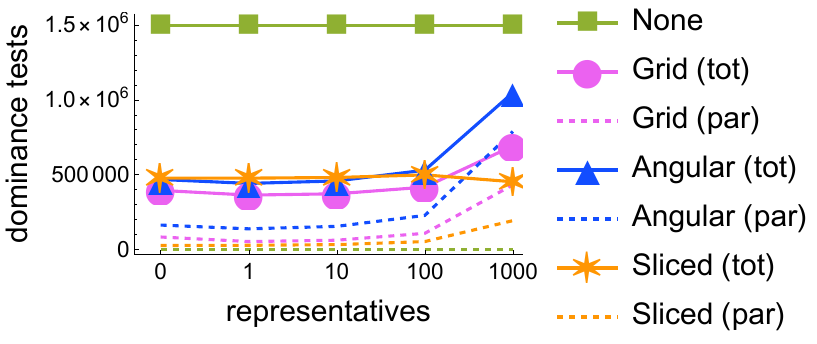}\label{fig:ant-varying-representatives}}%
\subfloat[][{\uniform}]
{\includegraphics[width=0.49\textwidth]{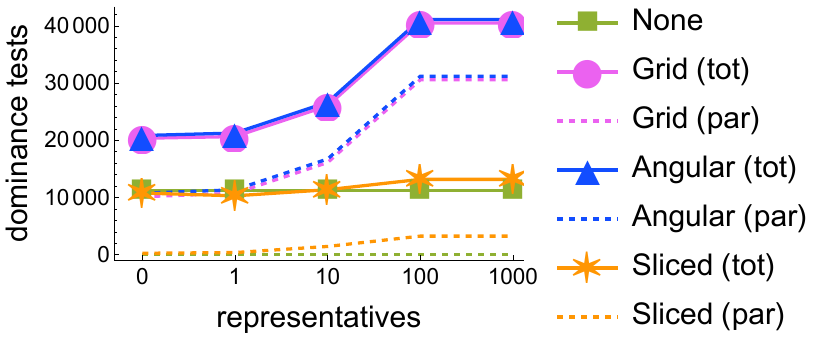}\label{fig:uni-varying-representatives}}%
\caption{Number of dominance tests with a default number of representatives ($\representatives=16$) as the number of partitions varies on \anticorrelated (\ref{fig:ant-varying-partitions}) and \uniform (\ref{fig:uni-varying-partitions}) 3D datasets with $\size=1$M tuples.}\label{fig:varying-representatives}
\end{figure}

\medskip

\noindent\textbf{Varying the number of representatives $\representatives$.}
We now measure the effects of the Representative Filtering technique on the computation of $\gridres$ by varying the number of representative tuples, $\representatives$, with default values for all other parameters for synthetic datasets.
Figure~\ref{fig:varying-representatives} clearly shows that, Representative Filtering is overall ineffective: in almost all considered scenarios, using representatives only overburdens the parallel phase with additional dominance tests without actually significantly reducing the union of local skylines.
Again, this is due to the fact that grid projections of skyline points are very likely to remain non-dominated, especially for smaller grid sizes, so that the pruning power of representative tuples fades.
The only case where a small advantage emerges is with \sliced on \anticorrelated with $\representatives=1000$, which determines a bare 5\% decrease in the number of dominance tests with respect to the case with no representatives.
We also observe that, to a target number of representatives corresponds a smaller average number of actually non-dominates tuples; for instance, on \anticorrelated, only $3.88$ tuples are non-dominated, on average, out of $10$ selected representatives, and only $132.36$ out of $1000$.

\medskip

\begin{figure}%
\centering
\subfloat[][{Smaller skylines}]
{\includegraphics[width=0.49\textwidth]{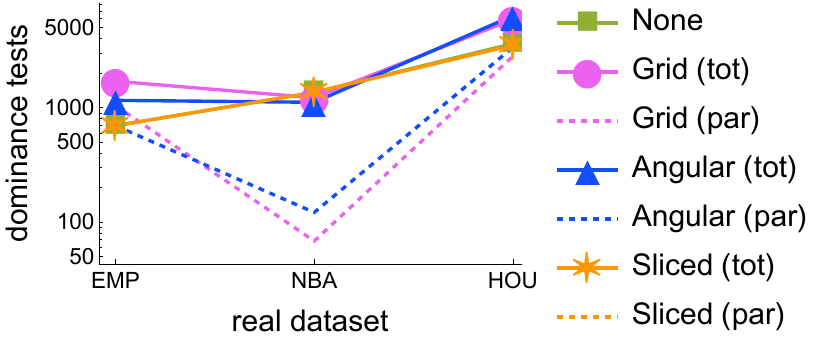}\label{fig:small-real}}%
\subfloat[][{Larger skylines}]
{\includegraphics[width=0.49\textwidth]{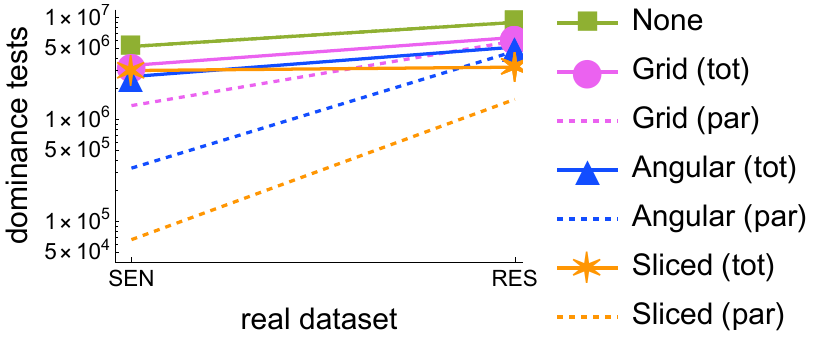}\label{fig:large-real}}%
\caption{Number of dominance tests with real datasets.}\label{fig:real-datasets}
\end{figure}

\noindent\textbf{Real datasets.}
Figure~\ref{fig:real-datasets} shows how the number of dominance tests varies depending on the partitioning strategy on different real datasets. Figure~\ref{fig:small-real} shows the three dataset with smaller skyline sizes: \emp, \nba, and \household, whose skyline sizes are, respectively, $14$, $14$, and $16$. Note that, while $\nba$ is a small dataset, the other two are much larger, but their tuples are correlated, thus causing a smaller skyline.
Figure~\ref{fig:large-real} shows what happens with real datasets with larger skylines: \sensors has $1496$ tuples in its skyline, while $\zillow$ has $8789$.
The results shown in the figure confirm what we found in the synthetic datasets: for the smaller cases, \angular and \grid do not pay off, while \sliced essentially has no parallel phase, thereby coinciding with or slightly improving over \none. With the larger datasets, the gains are at least 35\% with all strategies on \sensors and at least 29\% on \zillow, with peak improvements of 50\% on \sensors by \angular and 64\% on \zillow by \sliced.

\begin{figure}%
\centering
\subfloat[][{\anticorrelated}]
{\includegraphics[width=0.49\textwidth]{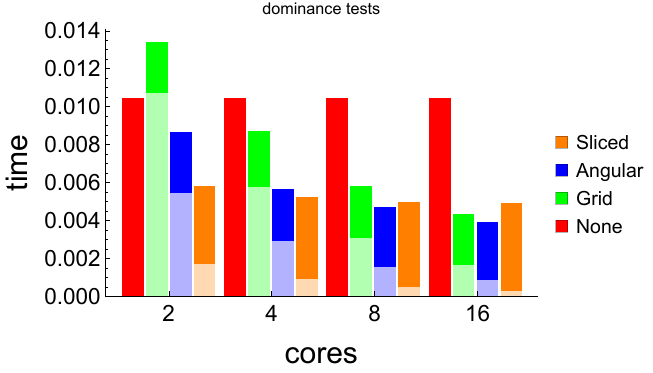}\label{fig:ant-varyingCores}}%
\subfloat[][{\zillow}]
{\includegraphics[width=0.49\textwidth]{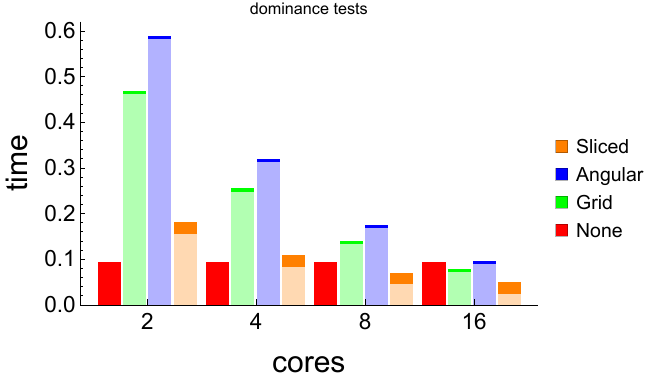}\label{fig:res-varyingCores}}%
\caption{Execution times on \anticorrelated (\ref{fig:ant-varyingCores}) and \zillow (\ref{fig:res-varyingCores}) as the number of cores varies.}\label{fig:execution-times}%
\end{figure}

\medskip

\noindent\textbf{Execution times.}
In order to get a sense of the concrete execution time required for completing the computation of $\gridres$ using the different partitioning strategies, we focus on \anticorrelated and \zillow, i.e., the most challenging synthetic and real datasets, respectively.
Figure~\ref{fig:execution-times} shows stacked bars reporting the breakdown of execution times for the various partitioning strategies with two components: the time needed to complete the parallel phase (lower part of the bar, shown in a lighter color) and the rest of the time (upper part), including the final, sequential phase and additional overhead caused by coordination of the execution over multiple cores.
In our execution setting, computing $\gridres$ for all the tuples in the skyline of \anticorrelated with default parameter values (Figure~\ref{fig:ant-varyingCores}) required $0.01$s with a plain, sequential implementation, in which the skyline is computed through a standard SFS algorithm~\cite{DBLP:conf/icde/ChomickiGGL03}.
Parallelization starts paying off already with just 2 cores with \angular and \sliced, while it takes at least 4 cores for \grid to surpass \none. With 16 cores, all strategies find the result in less than half the time required by \none.
We observe that, while all partitioning strategies improve as the number of cores grows, \sliced improves very little after 8 cores, since its parallel phase, for this dataset, is already very light when compared to the final phase.
Figure~\ref{fig:res-varyingCores} shows similar bars for the \zillow dataset, but here, due to the nature of the data, times are approximately ten times higher ($0.1$s for a plain sequential execution) and all parallel strategies incur high parallel costs, shown in the lower parts of the bars. While more cores are needed to perceive tangible improvements, all strategies attain a better performance than \none with 16 cores, all still having a large part of their execution time taken by the parallel phase, i.e., showing potential for further improving their performance with the availability of more cores (with 16 cores, \sliced already spends less than half the time taken by \none to compute the result).

\medskip

\noindent\textbf{Final observations.}
Our experiments show that the parallelization opportunities offered by the partitioning strategies we analyzed are useful for the computation of $\gridres$, provided that the application scenario is challenging enough to make the parallelization effort worthwhile, as we found it to be the case with the \anticorrelated datasets and with real datasets such as \zillow and \sensors. While there is no clear winner in all cases, \sliced provides the most stable performances across all datasets. The nature of the problem at hand, which deals with many tuples that are already strong, being part of the initial skyline, makes the Representative Filtering optimization ineffective and do not suggest to use an overly high number of partitions. While we conducted our analysis through a detailed counting of dominance tests, our results also indicate that a simple single-machine environment with 16 cores is sufficient to experience 2x improvements in the execution times in the most challenging datasets we adopted for our experiments.

\section{Related Work}
\label{sec:related}

In the last two and a half decades, the skyline operator has spurred numerous research efforts aiming to reduce its computational cost and augment its practicality by trying to overcome some of its most common limitations.

Skylines are the preference-agnostic counterpart of ranking (a.k.a. top-$k$) queries; while the former offer a wide overview of a dataset, the latter are more efficient, provide control over the output size, and apply to a large variety of queries, including complex joins and all sorts of utility components in their scoring function (i.e., the main tool for specifying preferences)~\cite{DBLP:journals/csur/IlyasBS08,DBLP:journals/pvldb/MartinenghiT10,DBLP:journals/tkde/MartinenghiT12}. Recently, hybrid approaches started to appear, trying to get the best of both worlds~\cite{DBLP:journals/pvldb/CiacciaM17,DBLP:conf/sigmod/MouratidisL021}.

As regards the efficiency aspects, several algorithms have been developed to address the centralized computation of skyline, including~\cite{DBLP:conf/icde/BorzsonyiKS01,DBLP:conf/icde/ChomickiGGL03,DBLP:journals/tods/PapadiasTFS05}.
In order to counter the most serious shortcomings of skylines, many different variants have been proposed so as to, e.g., accommodate user preferences and control or reduce the output size (which tends to grow uncontrollably in anticorrelated or highly dimensional data); a non-exhaustive list of works in this direction is~\cite{DBLP:journals/tods/CiacciaM20,DBLP:conf/cikm/CiacciaM18,DBLP:conf/sebd/CiacciaM18,DBLP:conf/sisap/BedoCMO19,DBLP:conf/sebd/CiacciaM19,DBLP:conf/sigmod/MouratidisL021}.

Improvements in efficiency have been studied also in the case of data distribution. In particular, \emph{vertical partitioning} has been studied extensively in the seminal works of Fagin~\cite{DBLP:conf/pods/Fagin98} and subsequent contributions, taming the so-called \emph{middleware scenario}. The strategies that we discussed here fall, instead, under the category of \emph{horizontal partitioning}, in which each partition is assigned a subset of the tuples. Peer-to-peer (P2P) architectures~\cite{DBLP:journals/tkde/CuiCXLSX09} first explored this approach, by having each peer to compute its skyline locally, and then to merge it with the rest of the network. With the maturity of parallel computation paradigms such as Map-Reduce and Spark, parallel solutions reflecting the pattern described at the beginning of Section~\ref{sec:parallel} have become common~\cite{DBLP:conf/edbt/MullesgaardPLZ14} and underwent careful experimental scrutiny~\cite{ciaccia2024optimizationstrategiesparallelcomputation}.

While skyline tuples are part of the commonly accepted semantics of ``potentially optimal'' tuples of a dataset, in their standard version they are returned to the final user as an unranked bunch. Recent attempts have been trying to counter this possibly overwhelming effect, particularly annoying in case of very large skylines, by equipping skyline tuples with additional numeric scores that can be used to rank them and focus on a restricted set thereof. The first proposal was to rank skyline tuples based on the count of tuples they dominate~\cite{DBLP:conf/sigmod/PapadiasTFS03}; albeit very simple to understand, subsequent literature has criticized this indicator for a number of reasons, among which the fact that 
\begin{enumerate*}[label=\emph{(\roman*)}]
\item it may be applied to non-skyline tuples, too, and thus the resulting ranking may not prioritize skyline tuples over non-skyline tuples,
\item too many ties would occur in such a ranking,
and 
\item it is not stable, i.e., it depends on the presence of ``junk'' (i.e., dominated) tuples.
\end{enumerate*}
Later attempts focused on other properties, including, e.g., the best rank a tuple might have in any ranking obtainable by using a ranking query with a \emph{linear} scoring function (i.e., the most common and, possibly, only type of scoring function adopted in practice)~\cite{DBLP:journals/pvldb/MouratidisZP15}. More recently, with the intention of exposing the inherent limitations of linear scoring functions, the authors of~\cite{CM:PACMMOD2024} introduced a number of novel indicators to measure both the ``robustness'' of a skyline tuple and the ``difficulty'' of retrieving it with a top-$k$ query. The indicators measuring difficulty are typically based on the construction of the \emph{convex hull} of a dataset, whose parallel computation has been studied extensively~\cite{DBLP:conf/pdcat/NakagawaMIN09,DBLP:conf/esa/WangYYD0S22,sym16121590}. Convex hull-based indicators include the mentioned best rank and the so-called \emph{concavity degree}, i.e., the amount of non-linearity required in the scoring function for a tuple to become part of the top-$k$ results of a query. As for robustness, the indicator called \emph{exclusive volume} refers to the measure of volume in the dominance region of a tuple that is not part of the dominance regions of any other tuples in the dataset; this indicator is computed as an instance of the so-called hypervolume contribution problem, which has also been studied extensively and is \#P-hard to solve exactly~\cite{DBLP:journals/csur/GuerreiroFP21,DBLP:journals/tcs/BringmannF12}. Finally, \emph{grid resistance} is the main indicator of robustness that we thoroughly analyzed in this paper.
The notion of \emph{stability}~\cite{DBLP:conf/sigmod/SolimanIMT11} is akin to robustness, in the sense that it still tries to measure how large perturbations can be tolerated to preserve the top-$k$ tuples of a ranking, although the focus is on attribute values in the scoring function and not on tuple values, as is done for grid resistance. To the best of our knowledge, apart from the sketchy sequential pattern given in the seminal paper~\cite{CM:PACMMOD2024}, there is no prior work trying to compute grid resistance, and even less so in a parallelized setting.

We also observe that both skylines and ranking queries are commonly included as typical parts of complex data preparation pipelines for subsequent processing based, e.g., on Machine Learning or Clustering algorithms~\cite{DBLP:conf/fqas/Masciari09,DBLP:journals/isci/MasciariMZ14}. In this respect, an approach similar to ours can be leveraged to improve data preparation and to assess the robustness of the data collected by heterogeneous sources like RFID~\cite{DBLP:conf/ideas/FazzingaFMF09,DBLP:journals/tods/FazzingaFFM13}.

\section{Conclusion}
\label{sec:conclusion}

In this paper, we tackled the problem of assigning and computing a value of strength to skyline tuples, so that these tuples can be ranked and selected accordingly. In particular, we have focused on a specific indicator of robustness, called grid resistance, that measures the amount of value quantization that can be tolerated by a given skyline tuple for it to continue to be part of the skyline.
Based on now consolidated algorithmic patterns that exploit data partitioning for the computation of skylines, we reviewed the main partitioning strategies that may be adopted in parallel environments (\grid{}, \angular{}, \sliced) as well as a common optimization strategy that can be used on top of that (Representative Filtering), and devised an algorithmic scheme that can be used to also compute grid resistance on a partitioned dataset.

We conducted an extensive experimental evaluation on a number of different real and synthetic datasets and studied the effect of several parameters (dataset size, number of dimensions, data distribution, number of partitions, and number of representative tuples) on the amount of dominance tests the are ultimately required to compute grid resistance. Our results showed that all partitioning strategies may be beneficial, with \grid often reaching lower levels of effectiveness than \angular and \sliced. We have observed that the specific problem at hand, in which one only manages skyline (i.e., inherently strong) tuples, makes Representative Filtering ineffective and suggests to not over-partition the dataset. Indeed, the relatively low value we used as default for the number of partitions ($\partitions=16$) also proved to be a good choice from the practical point of view. Our experiments on the execution times as the number of available cores varies confirmed the objective findings on the number of dominance tests and showed that, even with the limited parallelization opportunities offered by a single machine, the use of partitioning strategies may improve performances by more than 50\%, suggesting that there is leeway for further improvements with an increased number of available cores.

Future work will try to adapt or revisit the techniques used in this paper to the computation of other notions that are built on top of dominance. These include, e.g., skyline variants, based on modified notions of dominance, as well as other indicators of strength of skyline tuples, either novel or already proposed in the pertinent literature.

\end{document}